\documentclass[final]{ustcstep}
\usepackage{multirow,color}
\usepackage{bbding}
\usepackage{amssymb}
\usepackage{ulem}
\usepackage{color}
\usepackage[dvips]{graphicx}
\usepackage[square]{natbib}
\usepackage{abstract}
\usepackage{times}
\usepackage{multicol}

\def\gsim{\;\lower4pt\hbox{${\buildrel\displaystyle >\over\sim}$}\;}
\def\lsim{\;\lower4pt\hbox{${\buildrel\displaystyle <\over\sim}$}\;}
\def\grls{\;\lower4pt\hbox{${\buildrel\displaystyle >\over <}$}\;}

\newcommand\addr[2]{{\footnotesize \it $^{#1}$#2}\\}
\newcommand{\grad}{\ensuremath{^{\circ}}}
\def\grad{$^\circ$}
\def\kms{km s$^{-1}$}

\def\th{$^{th}$}

\begin{document}

\title{A Solar Coronal Jet Event Triggers A Coronal Mass Ejection}

\author{Jiajia Liu, Yuming Wang, Chenglong Shen, Kai Liu, Zonghao Pan, and S. Wang} 
\maketitle
\addr{}{CAS Key Laboratory of Geospace Environment, Earh and Space Science School, 
University of Science and Technology of China, NO. 96, JinZhai Road, Hefei, Anhui 230026, China}
\tableofcontents


\begin{abstract}
We present the multi-point and multi-wavelength observation and analysis on a solar coronal jet and coronal mass ejection (CME) 
event in this paper. Employing the GCS model, we obtained the real (three-dimensional) 
heliocentric distance and direction of the CME and found it propagate in a high speed over 1000 \kms{}. The jet 
erupted before and shared the same source region with the CME. The temporal and spacial relationship 
between them guide us the possibility that the jet triggered the CME and became its core. This scenario could promisingly enrich 
our understanding on the triggering mechanism of coronal mass ejections and their relations with coronal large-scale jets. 
On the other hand, the magnetic field configuration of the 
source region observed by the SDO/HMI instrument and the off-limb inverse Y-shaped configuration observed by SDO/AIA 171 \AA{} passband, 
together provide the first detailed observation on the three-dimensional reconnection process of large-scale jets as simulated in \textit{Pariat et al. 2009}. 
The erupting process of the jet highlights that filament-like materials are important during the eruption not only of small-scale X-ray jets 
\textit{(Sterling et al. 2015)} but also probably of large-scale EUV jets. Based on our observation and analysis, we propose a most possible 
mechanism for the whole event with a blob structure overlaying the three-dimensional structure of the jet to describe the interaction between the jet and the CME.

\end{abstract}

\section{Introduction}\label{sect:intr}

As one of the most intriguing phenomena acting in the solar atmosphere, solar jets have been 
studied extensively and deeply in the past few decades \citep[e.g.,][]{Shibata1996, Moore2010, Cirtain2007, DePontieu2007}. 
Despite the different properties in dominant temperature, scale and dynamics, they are thought to be import in releasing 
solar magnetic free energy through reconnection \citep[e.g.,][]{Shibata2007, Pariat2009, Liu2014, Fang2014} 
and contributing in corona heating and/or solar wind acceleration \citep[e.g.,][]{Shibata2007, Liu2014, McIntosh2010, 
Tsiropoula_Tziotziou2004, Liu2015}. As done by the numerical simulations in \cite{Pariat2009}, the triggering mechanism of 
solar jets could be credited to reconnection occurring within an inverse Y-shaped three-dimensional magnetic field 
configuration. However, direct observations on the detailed evolution of such a 3D reconnection are still absent. 

Playing impressive roles in affecting the Earth environment, coronal mass ejections (CMEs) have attracted great attentions 
since the era of space physics \citep[][as a review]{Low2001}. As a significant numerous matter ejected from the Sun, CMEs 
have been studied thoroughly in such as their observational features \citep[e.g.,][]{Wood1999}, models \citep[e.g.,][]{Priest1989, Lin2000}, 
early evolution \citep[e.g.,][]{LiuR2014}, interactions with each other \citep[e.g.,][]{Shen2012} and their arrival at the Earth \citep[e.g.,][]{Shen2014}. As widely 
believed in the community, a substantial part of CMEs are found to be associated with prominence/filament eruptions \citep[e.g.,][]{Gopalswamy2003, Gilbert2000}.

What hides behind the different typical geometries of jets (elongated) and CMEs (blob-like) is the distinct magnetic field configurations. While 
jets are usually believed to erupted along open field lines \citep[e.g.,][]{Shibata1996}, CMEs are thought to be associated with closed helical 
fields \citep[e.g.,][]{Chen1997}. Potential relations and interactions between these two glamorous phenomena (jets and CMEs) then will be intriguing and 
enrich our knowledge on the various physical processes in the solar atmosphere. Previous studies have shown that 
some (narrow) CMEs could be the extension of large-scale solar jets in white-light coronagraphs \citep[e.g.,][]{Wang1998, Liu2008}. 
Other possibilities might be one of them is triggered by the other. However, do these possibilities exist veritably in the solar 
atmosphere? How will such an interaction between a CME and a jet happen? Corresponding observations and analysis have not yet been performed.

In this paper, we will present the multi-point and multi-wavelength observation and analysis on a jet and CME event. The jet 
was originated from a single positive polarity active region, driven by a 3D magnetic topology which surprisingly resembled 
that in \cite{Pariat2009}, providing the first detailed observation on the evolution of such a 3D-reconnection triggering mechanism 
for solar jets. The CME is found to be triggered by the jet event with the jet becoming its core, illustrating a scenario in which these two 
eruptive events could be closely related. Conclusions come in Sect.~\ref{sect:conc} based on the observations in Sect.~\ref{sect:obs} 
and discussions in Sect.~\ref{sect:disc}.

\section{Observations}\label{sect:obs}
\begin{figure*}
\centering
\includegraphics[height=0.9\vsize]{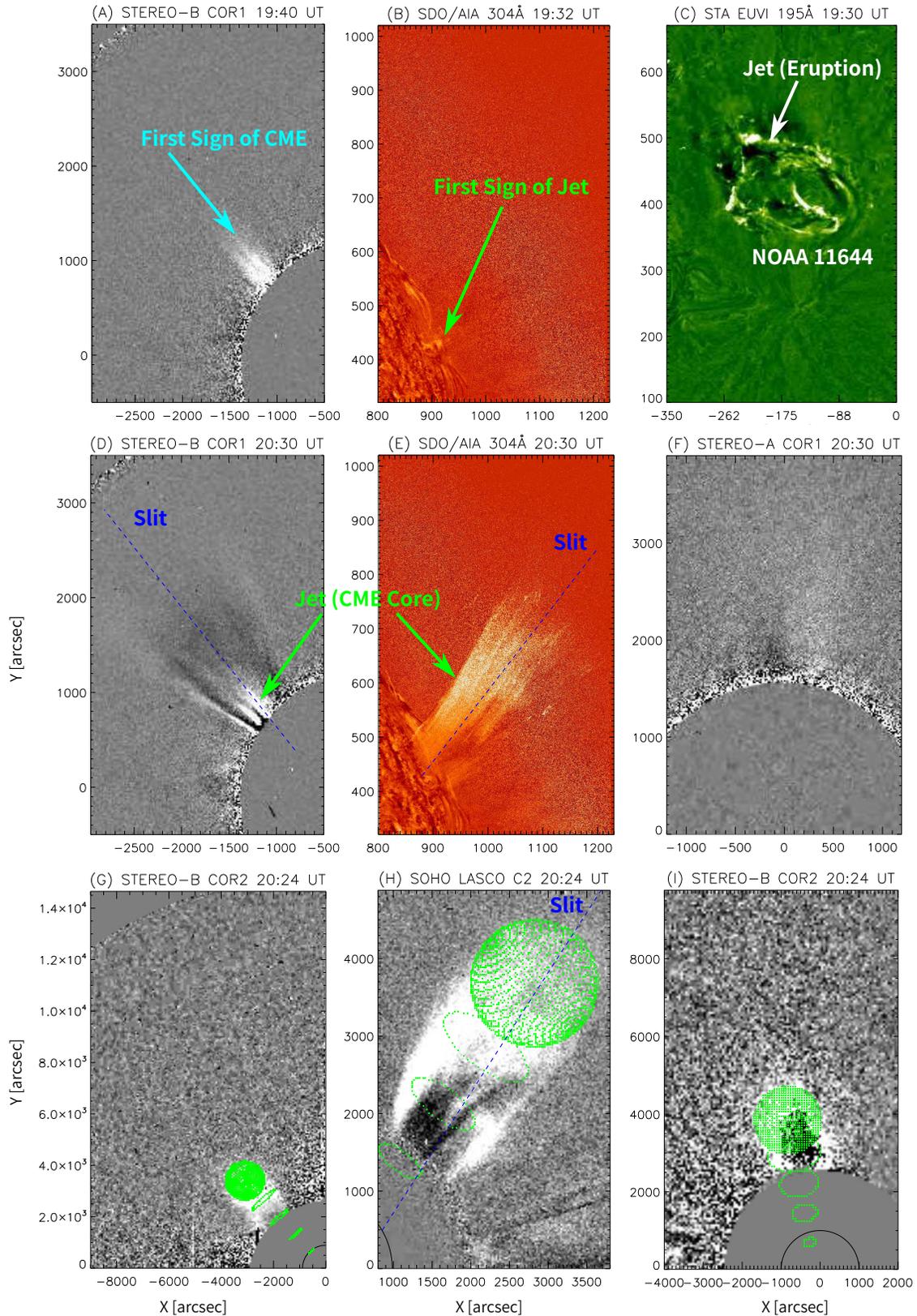}
\caption{Multi-point observations on the jet and the CME by SDO AIA, 
SOHO LASCO and STEREO EUVI/COR1/COR2 instruments. The top three panels are the 
STEREO-B COR1 observation on the first sign of the CME at 19:40 UT, the SDO AIA 
304 \AA{} observation on the first sign of the jet at 19:32 UT and the 
STEREO-A EUVI 195 \AA{} observation on the source region of the jet (active region NOAA 11644) 
at 19:30 UT, respectively. The medium three panels are the STEREO-A/B COR1 and 
SDO AIA 304 \AA{} passband observations at the same time 20:30 UT. The 
slits are selected along the axial direction of the jet or the CME. The bottom 
three panels are the simultaneous observations of the STEREO-A/B COR2 and SOHO 
LASCO C2 instruments on the CME on 20:24 UT. The green dotted wires 
are the resulting surface of the reconstured structure of the CME employing 
the GCS forward modeling method \citep[see descriptions about the model in the text and more in][]{Thernisien2009}.}\label{OV}
\end{figure*}

During the (quasi-) frontal travel of the active region NOAA 11644 (Fig.~\ref{OV} (C)), it performed several attractive 
eruptions, among which the most intriguing one is what acted after 19:00 UT on the day January 15\th\ 2013. A fast 
Coronal Mass Ejection (CME) with speed over 1000 kilometers per second could be observed simultaneously by 
the coronagraphs LASCO C2/C3 onboard the SOlar Heliospheric Observatory (SOHO) and COR1/COR2 onboard both probes of the 
Solar TErrestrial RElations Observatory \citep[STEREO,][]{Kaiser2008} (Fig.~\ref{OV}, online movie M1). Just before the CME, a solar 
coronal EUV jet was observed in the field-of-view (FOV) of the AIA instrument onboard 
the Solar Dynamics Observatory \citep[SDO,][]{Pesnell2012} and EUVI instruments onboard both of the STEREO probes (Fig.~\ref{OV}, online movie M2). 
Figure~\ref{OV} (B) and (C), together with the online movie M2, show the great simultaneity between the observed jet via 
SDO AIA and the eruption activities within the active region NOAA 11644 via STEREO-A EUVI, indicating that the active 
region should be the source region of the jet. Location of the center of the active region was about E10\grad{}N24\grad{} 
in the FOV of the STEREO-A EUVI instrument. As STEREO-A was about 129\grad{} ahead the earth (and SDO and SOHO) at 19:00 UT, 
the longitude of the center of the active region seen from SDO should be W119\grad{}.

The CME appeared at the north-east limb around 19:40 UT in the FOV of the STEREO-B (STB) COR1 (Fig.~\ref{OV} (A)), which was about 
134\grad\ behind the earth (and SOHO and SDO). Coronagraph LASCO C2 captured the first image of the CME at around 20:00 UT and half an hour 
later LASCO C3 and STEREO COR2. To obtain the three-dimensional (real) direction, position and velocity of the 
CME, we employ a forward reconstruction model (the GCS model) proposed by \cite{Thernisien2009}, which assumes a self-similar expanding 
flux rope structure of the target CME. We then compare the resulting flux rope obtained from the GCS model with the images from 
LASCO C2/C3 and STA/STB COR2 at six instances, when we could at least partially identify the leading edge of the CME in all the 
three images (LASCO C2, STA COR2 and STB COR2), to make the best agreement between the flux rope and the CME. 
Parameters of the flux rope could tell us the real heights (heliocentric distances), longitudes and latitudes of the CME 
at different times. One of them at 20:30 UT is shown in panel (G)-(I) in Figure~\ref{OV}, with the green dotted wires the surface 
of the flux rope structure.

The results display that the latitudes and longitudes of the CME stayed almost invariable around N48\grad\ and W120\grad\ 
seen from the Earth (and SDO) throughout its travel from 6 Rs to 20 Rs. A parabolic fitting between the heliocentric distance of the 
CME's leading edge and the corresponding time shows an average propagating speed of the CME around 1031 \kms{}.

On the other hand, continuous small brightenings could be figured out since 19:16 UT in the AIA 304 \AA{} passband images around 
N24\grad{} at the north-west limb and the first sign of plasma eruption which formed a jet appeared around 19:32 UT 
(Fig.~\ref{OV} (B), online movie M2 \& M3), few minutes earlier than when the CME appeared in the FOV at 
the STEREO COR1 observations (19:40 UT). Most of the jet materials were seen in the AIA 304 \AA{} passband and 
rarely found in passbands with characteristic temperature above 2 MK \citep[211 \AA{},][]{Lemen2012}, indicating that the jet was 
majorly formed with chromospheric cool materials (online move M2). The jet kept rising along a trajectory which was 
about 28\grad{} counter-clockwise away from the local radial direction and didn't show any signal of falling back. The position angle 
of the jet was almost the same with that of the CME (online movie M1)

\begin{figure*}[]
\begin{center}
\includegraphics[width=0.8\hsize]{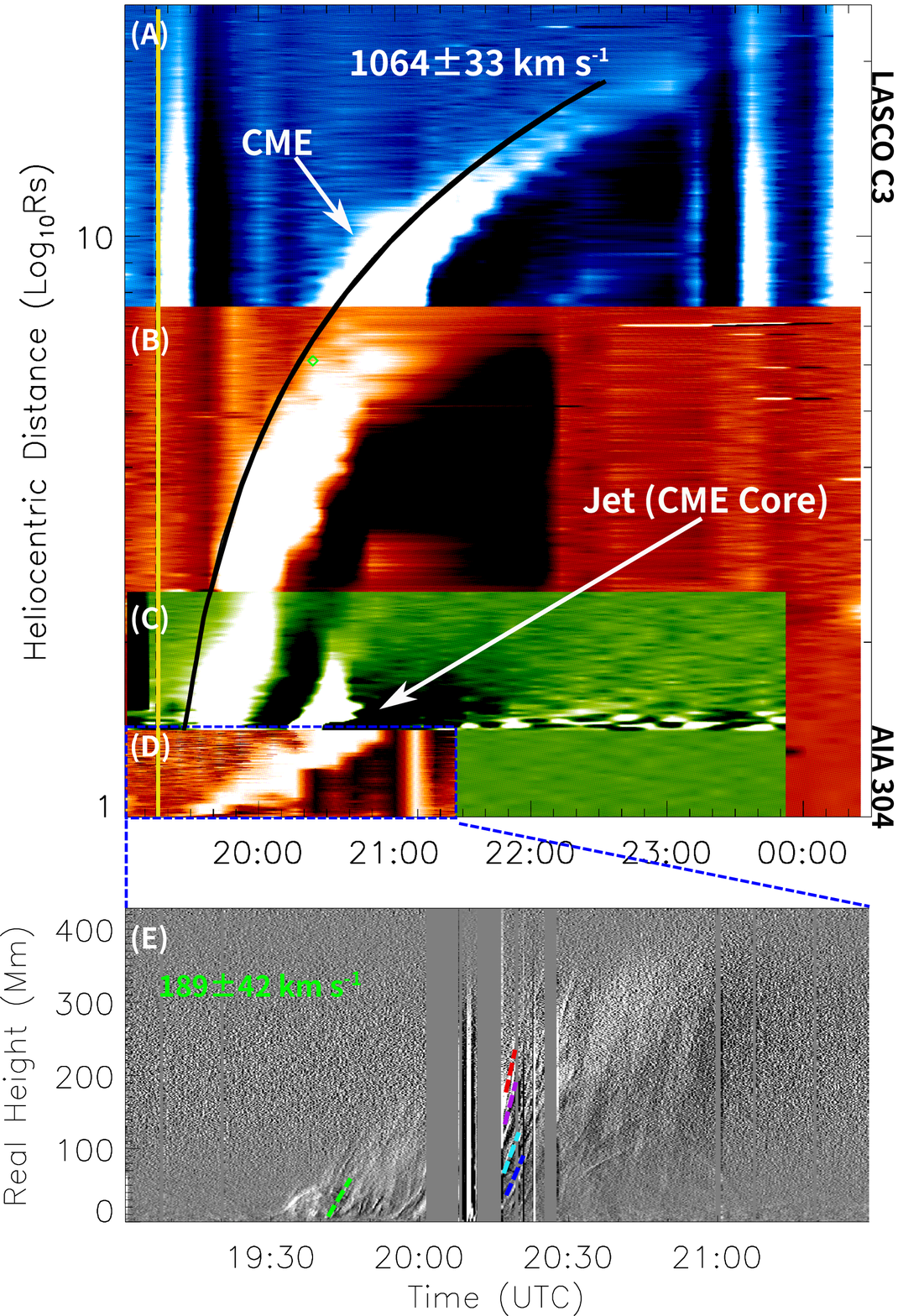}
\caption{Panel (A)-(C): Running-difference time-distance plots of the slits 
shown in Fig.~\ref{OV} based on data from SOHO LASCO C3, C2 and STEREO-B COR1 
respectively. Panel (D)-(E): Running-difference time-distance plots of the 
cadence-reduced and original SDO AIA 304 \AA{} passband data, respectively. 
The yellow vertical line across panel (A) to (D) indicates the time when the first 
brightening occurred at 19:16 UT inside the source region of the jet. The black 
solid curve across panel (A), (B) and (C) shows the 
fitted leading edge of the CME with a linear fit which results in an average 
speed of 1064 \kms{}. The green diamod shown in panel (B) is the leading edge 
reconstructed via the GCS model at 20:24 UT in panel (D)-(F) in Fig.~\ref{OV}. 
Color-coded lines in panel (E) give examples of tracks and velocities of different 
sub-jets.}\label{TD}
\end{center}
\end{figure*}

The similarity of the position and the temporal relationship between the jet and the CME, 
suggest that there should be some close relations between them: (1) the CME was the outer 
manifestation of the jet in the coronagraphs, (2) the CME was a bulk of materials 
that was triggered by the jet event, and (3) the lifting of the CME blob triggered footpoint region activities 
and the eruption of the jet. The third one could be directly excluded because the activities
which triggered the jet were more than 20 min earlier than when the CME was observed in the FOV of
STB COR1 (19:16 UT vs. 19:40 UT). To figure out which situation among the first two the truth is, we placed a slit along the CME's 
direction in the STB COR1 images (Fig.~\ref{OV} (D)), a slit along the jet's axis in the AIA images 
(Fig.~\ref{OV} (E)) and a slit along the CME's direction in the LASCO C2/C3 images (Fig.~\ref{OV} (H)) which 
was the extension of the slit in the AIA images. The corresponding time-distance plots are shown in 
Figure~\ref{TD} with the projection effect corrected based on the GCS result.

Obviously, the CME could be found in all the three running-difference time-distance diagrams for STB COR1 
(Fig.~\ref{TD} (C)) and LASCO C2/C3 (Fig.~\ref{TD} (B) and (A)) and they show very high consistency. Fitting the 
leading edge of the CME in the time-distance plots of these coronagraphs via a linear function yields the average velocity of 
it about 1064$\pm$33 \kms{}, which is highly consistent with the GCS result (1031 \kms{}). Figure~\ref{TD} (D) shows 
the corresponding running-difference time-distance plot of the AIA 304 \AA{} images with a deduced cadence of 
10 min to make it comparable to the STB COR1 plot. It is shown that there was no clear manifestation of the 
CME in the AIA 304 \AA{} observations (nor in the other EUV passbands of AIA which is not shown here, see movie M2). 
And the time sequence shows that the CME was not the extension of the jet in the coronagraphs.

High-cadence (12 s) AIA images could hint us some detailed kinetics of the jet. Figure~\ref{TD} (E) 
shows the de-projected running-difference time-distance plot along the slit in Figure~\ref{OV} (E) using the 
12s-cadence images of the AIA 304 \AA{} passband. Sub-jets were expelled successively, indicating that 
continuous reconnection happened around the source region of the jet as described in \cite{Moore2010} and observed 
in \cite{Liu2014}. The average axial speed of these sub-jets was about 189$\pm$42 \kms{} at the bottom (green dashed 
line in Fig.~\ref{TD} (E)) and they could be found to undergo obvious acceleration. However, it is difficult for 
us to trace the track of a particular sub-jet in the time-distance plot to estimate the exact acceleration, 
due to the 20-min data gap from about 20:01 UT. The speed of the jet in the FOV of STB COR1 was found to be almost the 
same with that of the CME's leading edge, together with the great 
continuity between the AIA plot and coronagraph plots, indicates the second situation that the CME was triggered by 
the jet event with the jet becoming its core should reveal the truth.

\section{Mechanisms and Discussions}\label{sect:disc}

\begin{figure*}[tbh!]
\begin{center}
\includegraphics[width=0.9\hsize]{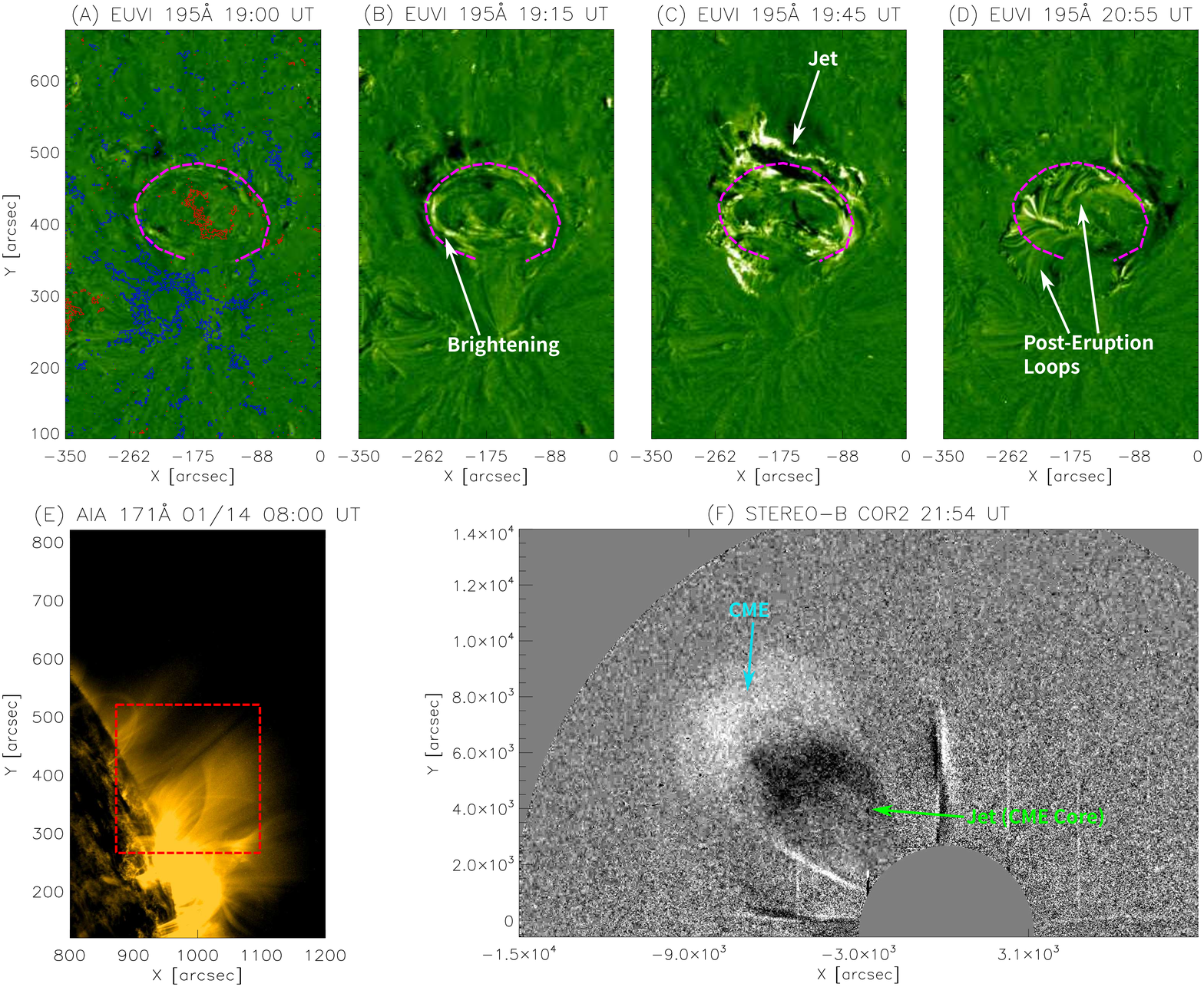}
\caption{Panel (A)-(D): Time-sequence evolution of the source region 
(NOAA 11644) of the jet and the CME in the FOV of STEREO-A EUVI 
195 \AA{} passband. The red and blue contours in (A) show the 
positive and negative magnetic fields observed by SDO HMI instrument 
nine days before the jet and CME event. The purple dashed curve marks 
the filament-like materials laying around the polarity inversion region 
before the eruption. Panel (E) is the SDO AIA 171 \AA{} image 
at 08:00 UT one day before the event, resolving an inverse Y-shaped 
configuration of the magnetic field lines enclosed in the red dashed 
rectangle. Panel (F): STEREO-B COR2 observation on the CME and the jet 
(CME core) at 21:54 UT.}\label{RE}
\end{center}
\end{figure*}

The source region of the jet which was identified as the active region NOAA 11644 could be seen clearly from the 
STEREO-A EUVI 195 \AA{} observations. Figure~\ref{RE} (A)-(D) exhibit four different times before, during and 
after the eruption. Figure~\ref{RE} (A) shows the active region before the eruption at 19:00 UT, with the red-blue 
contours the line-of-sight (LOS) photospheric magnetic fields observed by the HMI instrument onboard SDO at 
17:00 UT nine days earlier when the active region was almost facing the satellite. It is shown clearly that the 
active region contained a single positive polarity surrounded by negative polarities, with another positive 
polarity at the lower left in a southern active region. Filament-like materials marked as a purple dashed curve could 
be found laying around the polarity inversion region between the single positive polarity and the surrounding 
negatives. Brightenings began around 19:15 UT as shown in panel (B) and developed into the eruption of the jet 
which was seemingly formed by the laying filament-like materials. The eruption began around the left end of the purple 
curve, gradually proceeded clockwise and rooted around the right end, as shown in panel (C). Newly formed loops 
could also be notably observed after the eruption of the jet, exhibited in panel (D). As we can find from Movie 2 and Figure~\ref{RE}, 
the filament-like materials participated a lot during the triggering process and finally formed the erupted jet.

Figure~\ref{RE} (E) shows the AIA 171 \AA{} observation at 08:00 UT one day before the event and not far after 
the active region turned to the limb. Images in the 171 \AA{} passband can mostly resolve the magnetic field 
topology in the lower corona and showed a clear inverse Y-shaped configuration (enclosed in the red dashed 
rectangle in Fig.~\ref{RE} (E)), which almost kept exact pace with the simulations of solar jets in \cite{Pariat2009} and seemed 
like a magnified ``anemone" jet's root observed in \cite{Shibata2007}. The photospheric magnetic field, the 
off-limb inverse Y-shaped configuration, the erupting process of the jet's materials and the existence of newly 
formed loops after the eruption are greatly analogous to the 3D-reconnection simulations in \cite{Pariat2009}, 
inspiring us to surmise similar three-dimensional reconnections happening in this particular event. 

\begin{figure}[t!]
\begin{center}
\includegraphics[width=0.9\hsize]{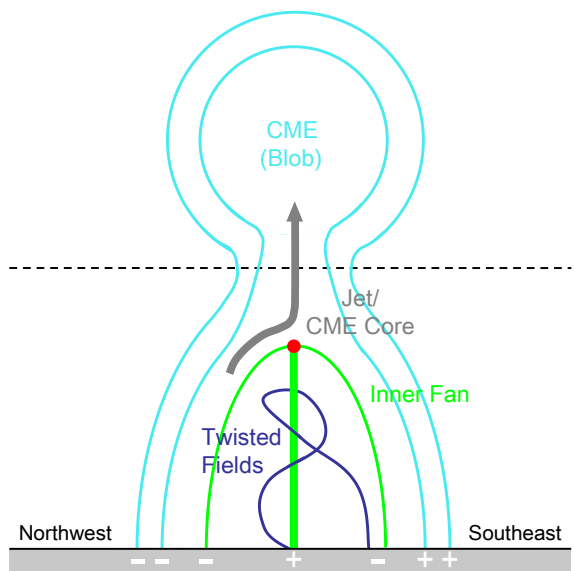}
\caption{2D-sketch vesion of the magnetic field configuration 
that results in the observed event and phenomena. See the text 
for details of the illustration on the mechanisms.}\label{MD}
\end{center}
\end{figure}

Based on the above observation and analysis, we are then able to propose the most possible mechanism for the whole event, which is 
shown as a 2D-version sketch in Figure~\ref{MD}. Periphery field lines that originate from the single positive 
polarity inside the active region 11644 and end at the negative polarities form the inner fan (green fields in Fig.~\ref{MD}). 
Out of which are the bottom fields of the CME (cyan fields) originating from the southern positive polarity. 
If activities underneath the inner fields introduce any twist/shearing (shown as the blue field), magnetic free 
energy will be built up and reconnection will occur when the balance is broken up \citep{Pariat2009, Fang2014}. 
Plasma materials speed up by the reconnection will then form a jet (gray-colored arrow). The bursts could also 
give a push to the blob structure upon the black dashed horizontal 
line. Due to the lack of observations on the early stage of the blob structure, we cannot know how the bursts 
pushed it in this particular event. Several effects such as: (1) the elevated inner fan due to 
the reconnections \citep{Pariat2009} or/and (2) the accelerated jet materials could lead to the ascending of the blob. 
Under the effect of one or more of the above processes, the blob rises and forms the observed CME.

Like the the one modeled in \cite{Fang2014} and the one observed in \cite{Liu2014}, rotational motions of 
the jet's material could also be observed from the AIA 304 \AA{} images in this event. Placing slits perpendicular 
to the jet's axis (which are not shown in the figures) allow us to estimate the rotating periods of the materials 
employing a sine-function fit as done in \cite{Liu2014}. Periods turn out to be about 15 min at the bottom with 
the resulting linear speed of about 280 \kms{}. The periods became longer at 
the top of the jet in the FOV of AIA, indicating a deceleration of the rotational motion. As illustrated in \cite{Liu2014}, 
the rotational motion comes with the release of residual magnetic free energy after the reconnection, how much the 
residual is would affect the rotational motion of the jet. Thus it is possible that in this particular case, 
the reconnection may have already released most of the magnetic free energy and few of them is left to drive the 
rotational motion.

However, another situation that the angular momentum of the jet was passed into that of the CME is 
not prohibited. As lack of direct observations on the rotational motion of the CME 
\citep[discussions on this particular issue could be found in][as an example]{Tian2013}, we could try to figure 
out if this situation is possible from the poloidal motion of magnetic clouds (MCs) which are the counterpart of CMEs in the 
interplanetary space. As derived from a velocity-modified cylindrical force-free flux rope model base on Wind 
observations, \cite{Wang2015} found that almost all the MCs more or less had shown poloidal motion with meridian 
linear speed around 10 \kms{}. Assuming all the angular momentum of the jet has been passed to the CME, a self-expanding 
propagation of the CME and the conservation of angular momentum in the interplanetary space, the poloidal speed 
of the MC, evolved from the CME in this particular case, would be about 1.3 \kms{}, which is slightly below but still
not prohibited by the observations in \cite{Wang2015}.

\section{Conclusions}\label{sect:conc}
In this paper, we present the analysis of a coronal jet and a related coronal mass ejection (CME), 
employing multi-wavelength and multi-point observations from SDO AIA/HMI, STEREO EUVI/COR1/COR2 and SOHO LASCO C2/C3 instruments. 

Detailed analysis has shown that there is a close relationship between the jet and the CME. Employing the GCS 
reconstruction model, we are able to obtain the real heights and positions of the CME at different times. It is found that the 
CME propagated with a speed over 1000 \kms{}. After correcting the projection effect using results from the GCS model, 
we plot the time-distance diagrams of the CME in the FOV of STB/COR1 and LASCO C2/C3, and of the jet in the FOV 
of SDO/AIA. It is shown that the CME was not the extension of the jet in coronagraphs, but triggered by the jet 
event, with the jet becoming its core. All the observations indicate that a (high-speed) CME could also be triggered by an 
underneath jet event, which is quite different with classical models \citep[e.g.,][]{Lin2000} and provide a new viewpoint on 
studying the relations between these two different mass release events in the solar atmosphere. 

The jet originated from a source region with single positive polarity surrounded by 
negative polarities. All the observational features and its source region configuration showed highly consistence with the three-dimensional 
reconnection simulation of solar jets in \cite{Pariat2009}, providing the first detailed observations on such three-dimensional reconnection 
triggering process of large-scale EUV jets. The erupting process of the jet observed by the STEREO-A 194 \AA{} instrument highlights the 
importance of the footpoint region filament-like materials in participating the eruption, which might provide another evidence on the model 
proposed by \cite{Sterling2015} (in which the authors showed evidences for X-ray small-scale jets).

Rotational motion of the jet materials could also be observed in the AIA 304 \AA{} images. The rotational 
motion was found to decelerate. The deceleration could be either caused by that: (1) little magnetic free energy was left 
after the reconnection, or/and (2) the angular momentum of the jet had been passed to the CME. As lack of 
observations on the temporal magnetic fields, early stages of the CME and in-situ data, we cannot exclude either 
one of these two explanations. On the other hand, as the CME was invisible in the observations of the AIA 
instrument due to the different temperature or/and the submarginal height, we are not able to figure out 
how exactly the CME interacted with the underneath jet event. This made the physical image incomplete 
in describing the interactions between the CME and the jet event of our picture in Figure~\ref{MD}. 
Future works on more observations and numerical simulations may shed light on the issues raised above.

\acknowledgments{We acknowledge the use of data from the Solar Dynamics Observatory (SDO), the Solar TErrestrial RElations 
Observatory (STEREO) and the SOlar Heliospheric Observatory (SOHO). Online movie M1 is generated using 
the ``JHelioviewer" tool (http://www.helioviewer.org). This work is supported by grants from 
NSFC (41131065, 41121003 and 41574165), CAS (Key Research Program KZZD-EW-01-4), MOST 973 key project (2011CB811403), 
MOEC (20113402110001) and the fundamental research funds for the central universities. The research leading to 
these results has also received funding from NSFC 41274173, 41404134 and 41304145.}

\bibliographystyle{agufull}

\end{document}